\documentstyle[twocolumn,prl,aps]{revtex}  %,12pt

\begin{document}
\draft

\title{\bf Probability of a Solution to the Solar Neutrino Problem
Within the Minimal Standard Model\\
       }  %\normalsize
\vspace{.25in}
\author{\vspace{.25in} {\bf Karsten M. Heeger and R.G.H. Robertson} }
\address{\em Nuclear Physics Laboratory,
        University of Washington,  Seattle, WA 98195, U.S.A.}

%\finalcopy

\maketitle
\date{\today}

\setlength{\baselineskip}{3.2ex}
\vspace{.5in}

\begin{abstract}
Tests, independent of any solar model, can be
made of whether solar neutrino experiments are consistent with the minimal
Standard Model (stable, massless neutrinos).   If the experimental
uncertainties are correctly estimated and the sun is generating energy by
light-element fusion in quasi-static equilibrium, the probability of a
standard-physics solution is less than 2\%.  Even when the luminosity
constraint is abandoned, the probability is not more than 4\%.  The sensitivity
of the conclusions to input parameters is explored.
\end{abstract}

%\vskip-1pc


\narrowtext

%\section{Introduction}
%\label{sec:intro}

The sun is believed to generate its energy by fusion reactions that can be
summarized as
	$$4 {\rm p} + 2{\rm e}^- \rightarrow \ {^4{\rm He}}  +  2\nu_e + {\rm 26.731
\ MeV.}$$
A number of pathways lead to $^4$He, and a complex spectrum of
neutrinos from  $pp$, $pep$, $^7$Be, $hep$, $^{13}$N, $^{15}$O,
$^{17}$F, and
$^8$B results \cite{bahcallbook}.  The spectral shape of each individual
component, whether line or continuum, is determined by laboratory measurement
and/or electroweak theory.  Its relative intensity, on the other hand, depends
delicately on astrophysical models of the sun.     The fact that these models
predicted to within a factor of two the intensity of a 0.01\% branch
($^8$B) that
varies as the 25th power
\cite{BacUlmer} of the central temperature of the sun must be regarded as a
stunning achievement and a clear indication of the basic correctness of our
understanding of how the sun and other stars function.

Nonetheless, the lack of perfect agreement raised speculation about possible
exotic origins, such as neutrino oscillations.  At first, the model
dependence of
the $^8$B flux calculation made such speculations interesting but not
compelling.  Now, however,  steadily improving data from 4 independent
experiments are available.  The Homestake Cl-Ar experiment \cite{Davis} gives 	
2.55
$\pm$ 0.17 $\pm$ 0.18	 solar neutrino units (SNU),  and the  Kamiokande
\cite{Kam3} result (increased 2\% by radiative corrections \cite{Bahcallrad})
is 	(2.95
$^{+0.22}_{-0.21}\ \pm$ 0.36)	x 10$^6\ ^8$B $\nu_e$ cm$^{-2}$ s$^{-1}$.  For the
SAGE
\cite{Sage} and Gallex \cite{Gallex} experiments, a
weighted average of  73.8
$\pm$ 7.8 SNU is adopted \cite{FN1}. (1 SNU =   $10^{-36}$ events per atom
per second.)

%\section{Calculation of Spectrum}

Because the three types of experiment have different energy thresholds, a coarse
neutrino spectroscopy of the sun has been made.  The least model-dependent
questions that can be asked are,

 {\em Is it possible to describe the neutrino spectrum with any combination of
the known sources in hydrogen-burning?

Is the total neutrino flux consistent with the solar luminosity?}

Many have considered model-independent analyses
\cite{Hata,Bahc4,Parke,Shi,Hata0,Hata2,Gates,Ber,Castellani}; in particular,
Hata {\em et al.} \cite{Hata0} showed the data to be inconsistent with hydrogen
burning and the luminosity constraint without new physics. To this body of
analysis we add (a) a
test of consistency free of the luminosity constraint, (b) a test for
inconsistency of the data with the total solar luminosity, (c) the probabilities
that the existing data would be obtained from true values in the physical regime
in the absence of new physics, and (d) the dependence of the
conclusions on the neutrino cross sections.

The spectral shape and endpoint of the neutrino data from Kamiokande show that
$^8$B neutrinos are emitted from the sun and that $hep$ neutrinos are, as
expected, negligible.    The $pep$ reaction rate we take to be a fixed fraction,
$f_{pep} = 0.23(2)$\%,  of the $pp$ rate, \cite{BP,Bahc3} (while in
principle model-dependent, $f_{pep}$ is one of the most
reliably determined model parameters, depending chiefly on the electron density
and only weakly on temperature and on nuclear wavefunctions
\cite{bahcallbook}).  The
$^7$Be and CNO fluxes play a qualitatively interchangeable role in the existing
experiments -- the Cl-Ar and Ga experiments are sensitive to both and
Kamiokande to neither.  As a result, it is possible to draw very general
conclusions without knowledge of the relative sizes of each.

Defining the $pp + pep$, $^7$Be + CNO, and $^8$B fluxes as $\Phi_1$,
$\Phi_{7+}$,
and
$\Phi_8$, respectively, the experimental capture rates  as $R_{Cl}$ for
Cl-Ar and $R_{Ga}$ for Ga-Ge, and the
experimental  $^8$B flux from Kamiokande as $R_{Kam}$, the following equations
result:
 \begin{eqnarray}
a_{C1} \Phi_1 + a_{C7} \Phi_{7+} + a_{C8} \Phi_8 & = & R_{Cl} \\
                                   a_{K8} \Phi_8 & = & R_{Kam} \\
a_{G1} \Phi_1 + a_{G7} \Phi_{7+} + a_{G8} \Phi_8 & = & R_{Ga}
\end{eqnarray}

%\vfill\eject

 The coefficients, with the neutrino physics of the minimal Standard Model
(MSM),  are listed in Table
\ref{coeffs}. The parameter $f_{CNO}$ is the fraction of the flux  $\Phi_{7+}$
that is due to CNO reactions (0$\leq f_{CNO} \leq$1). In Table \ref{fluxes} are
shown the values of the fluxes obtained by propagating the uncertainties in the
cross sections and solving. One finds that $\Phi_{7+}$ is {\em always} negative,
at the same confidence level, irrespective of the value of
$f_{CNO}$.

A negative flux is unphysical. Remarkably, the initial
premise that the data can be described as the sum of $pp + pep$, $^{7}$Be,
CNO, and $^8$B electron-neutrino spectra, in any proportions whatsoever,
fails at
the 96\% confidence level \cite{PDG}.

A fourth neutrino-flux relationship is contained in the total solar luminosity,
for a quasi-static sun deriving its energy entirely
from hydrogen burning.  When neutrino losses are accounted for, the
electromagnetic solar constant (irradiance) $I$  in 10$^{10}$ MeV cm$^{-2}$
s$^{-1}$ is given by:
 \begin{eqnarray}
0.980 (1 - 0.088f_{pep})\Phi_1
+ 0.939 (1 - 0.003f_{CNO}) \Phi_{7+}  \nonumber \\
+ 0.498\Phi_8     = \frac{2I}{Q}
\end{eqnarray}
Experimentally, $I$ = 85.31(34)  \cite{BP},  and Q = 26.731 MeV. Additional
flux constraints for hydrogen burning are given by Bahcall and Krastev
\cite{BK}.

Under the assumption of hydrogen
burning,  Eq. (1-3) can be recast with variables $I$, $\Phi_{7+}$, and
$\Phi_{8}$
(for example). The irradiance is found to be 101(18), in agreement with the
experimental value,  but, as before,
$\Phi_{7+}$ = -0.43(24) x 10$^{10}$ cm$^{-2}$ s$^{-1}$.
On the other hand, forcing  $\Phi_{7+}$ to zero yields $I$ =  72(8), and
$\chi^2$
= 3.2 for 1 degree of freedom.  (Principally it is the gallium experiments that
induce the strong negative correlation between the irradiance and
$\Phi_{7+}$.)  Thus, while any MSM solution is relatively improbable, the solar
neutrino problem is not necessarily manifest in the total neutrino flux.

Including the photometrically measured luminosity as a fourth constraint
reduces the uncertainties in the derived fluxes, as summarized in Table
\ref{fluxes}. The probability of this result being obtained from a
physically-realizable set of fluxes ({\em i.e.} with the $^7$Be + CNO flux being
non-negative) is  less than 2\%, and quantifies directly, for example, the
``last hope'' suggested by Berezinsky et al. \cite{Ber2}.

The luminosity constraint, Eq. (4), defines a plane in $\Phi_1  \Phi_{7+}
\Phi_8$ space.  Solutions allowed in the MSM {\em must} fall
within the triangular region of this plane in the positive octant (Fig.
\ref{4par}).  The data do
not meet this condition.

 The assumptions made in reaching this conclusion do not include any
features of
solar models  (one  \cite{BP} is
shown, for reference, in Table \ref{fluxes}). Therefore,
the shape of the $^8$B spectrum is not as expected, containing more strength at
high energies and less at low \cite{deB}, and/or the neutrino flavor
content is not pure electron, which alters the relationship between the
Kamiokande result and the radiochemical experiments (because Kamiokande detects,
via the neutral-current interaction,  neutrinos of all active flavors).  These
features are characteristic of neutrino-oscillation solutions
\cite{Hata,Bludman,Krastev,Fiorentini}.  In contrast to the standard-physics
solution, such solutions give an excellent account of all data.  Once such
solutions are admitted, the fluxes may in general be quite different
\cite{Rosen,BahcallCNO}.

%\section{Sensitivity to Inputs}

While no astrophysical model inputs have been used in the analysis, the
conclusions do depend on both neutrino cross sections and experimental
uncertainties (statistical and systematic).  The dependences serve to highlight
the most critical experimental inputs, and aid in planning  future
experimental work.  In Table \ref{diff} the differential coefficients for the
$^7$Be + CNO flux
$\Phi_{7+}$ are tabulated.

  Although it is a common perception
that the solar neutrino problem stands or falls on the validity of the Cl-Ar
experiment, the Kamiokande datum is twice as critical.  By `critical' is meant
the number of standard deviations change in an experimental result to produce a
given change in $\Phi_{7+}$, {\em i.e.} the value of
$\frac{\partial \Phi_{7+}}{\partial  R}\sigma_R $.

The Ga data are almost irrelevant in the determination
of the
$^7$Be + CNO flux when the luminosity is a free parameter, but
 dominate it when the luminosity is input. This sensitivity draws attention to
the importance of the neutrino cross sections, $a_{G7}$ and
$a_{G8}$, which are determined in part by (p,n) reactions to excited states,
with uncertainties that are difficult to assess.  Hata and Haxton \cite{Haxton}
have pointed out that the Gallex  \cite{Calib} and SAGE \cite{Sage} $^{51}$Cr
source calibration experiments are, in fact,  experimental confirmation that
$a_{G7}$ is close to the expected value unless a novel effect has caused the
extraction efficiency to be low, and $a_{G7}$ is correspondingly larger than
expected. In the latter case, the calibration data make the detector response to
$^7$Be neutrinos virtually independent of the efficiency, while the response to
$pp$ and
$^8$B neutrinos scales linearly with the efficiency. The efficiency of
Gallex and SAGE would both have to be reduced to 77\% of the measured values to
bring the derived  $^7$Be + CNO flux up to zero, at which point $\chi^2$
exceeds 4.

Could the present situation  reflect an
experimental result outside  its estimated uncertainty? Luminosity-constrained
fits of the three types of experiment in pairs give for $\Phi_{7+}$ the
following values (10$^{10}$ cm$^{-2}$ s$^{-1}$): Kamiokande--Cl-Ar, -0.39(22);
Gallium--Cl-Ar, -0.18(12); Kamiokande-Gallium, -0.19(11).  The anomaly emerges
from all combinations of pairs of experiment.  This fact has the corollary
that, since Gallium and Cl-Ar have no neutral-current sensitivity, a
non-standard  $^8$B spectrum shape is somewhat favored.  Experimental
uncertainties in this shape contribute about 2\% \cite{Bahc6} to the error in
$a_{C8}$ and somewhat more \cite{deB} to that in $R_{Kam}$,  but in a correlated
way that diminishes the effect on $\Phi_{7+}$. New laboratory
determinations of the spectrum are highly desirable.

%\section{Conclusions}

At an interesting level of confidence  (about 98\%), there exists a solar
neutrino problem independent of solar models, except for the assumptions of
neutrino production by light elements and a steady-state sun. Moreover, even
abandoning the steady-state sun assumption (or, equivalently, postulating
exotic energy sources) does not deliver a satisfactory solution at the 96\%
confidence level. With unpublished new data \cite{nu96} these confidence levels
reach 99.5\% and 94\%, respectively.  The numbers quantify the minimum extent of
the problem in the sense that neither the  $^7$Be nor the CNO flux can
actually be
 {\em exactly} zero.    At the present level of significance,
the data suggest new neutrino physics, and, at the same level,  demonstrate that
the solution to the solar neutrino problem is not to be found in the realm of
astrophysics.  While we keenly await results from the new generation of
experiments \cite{dpf},  SuperKamiokande, SNO, and Borexino, we emphasize that,
in this approach, there is also much to be gained from improvements to existing
experiments.  To illustrate the potential, setting to zero the statistical
errors
in the present experiments gives a result incompatible with standard physics at
$>$99.998\% confidence level.  On the other hand,  systematic uncertainties are
notoriously difficult to estimate, and caution is advisable.  We also
underscore the value of experimental work on the important cross sections
$a_{C8}$, $a_{G7}$, and $a_{G8}$, and the shape of the $^8$B spectrum, in the
task of clarifying this fundamentally important question.

We have benefited greatly from discussions with  J.N. Bahcall, S. Brice, N.
Hata,  W. Haxton,  and J.F. Wilkerson.  A. Cumming kindly corrected a number of
errors.  This work was supported by the U.S. Department of Energy under
Grant No.
DE-FG06-90ER40537.

\vfill\eject

\begin{figure}
\caption{The luminosity plane defined by Eq. 1, and (inset) the
1.64-standard-deviation contours (95\% confidence level for $\Phi_{7+}$)
from the
data for selected values of
$f_{CNO}$. The fluxes are in units of 10$^{10}$ cm$^{-2}$ s$^{-1}$. Solutions
allowed by the MSM and the luminosity constraint must fall
within the triangular area.  Below the dashed lines parameterized by
$f_{CNO}$,  the Bahcall-Krastev condition  $\Phi_1 \geq
\Phi_{7} + \Phi_8$ required in hydrogen burning is not met.}
\label{4par}
\end{figure}

\vfill\eject

\begin{table}
\caption{Cross-section coefficients.}
\label{coeffs}
%\medskip
%\centering
\begin{tabular}{lllr}
&  	& Cross Section	&  Reference \\
& & 10$^{-46}$ cm$^2$ &  \\
\hline
& $a_{C1}$ & 16 $f_{pep}$  &    \protect{\cite{BP,BU}} \\
Cl-Ar & $a_{C7}$ & 2.38(1 + 2.60$f_{CNO}$)  &    \protect{\cite{Bahc5}} \\
& $a_{C8}$ & 11100  &   \protect{\cite{Trinder,Bahc6}} \\
\hline
Kamiokande & $a_{K8}$ & 10000$^a$  &    \protect{\cite{Bahcallrad}} \\
\hline
& $a_{G1}$ & 11.8(1 + 17$f_{pep}$)  &    \protect{\cite{BU,Hampel}} \\
Gallium & $a_{G7}$ & 76.5(1 + 1.42$f_{CNO}$)  &    \protect{\cite{Haxton}}
\\ & $a_{G8}$ & 24600  &    \protect{\cite{Bahc6}} \\
\hline
\multicolumn{3}{l}{$^a$Dimensionless. Kamiokande reports flux directly.}
\end{tabular}
\end{table}

\mediumtext

\begin{table}
\caption{Fitted values of the fluxes (10$^{10} \nu$ cm$^{-2}$ s$^{-1}$)
($f_{CNO} = 0$, $f_{pep} = 0.0023$). }
\label{fluxes}
%\medskip
%\centering
\begin{tabular}{llllll}
Component	& Value	& Uncertainty &  Value  & Uncertainty &  SSM$^a$ \\
& \multicolumn{2}{l}{(Luminosity unconstrained)} & \multicolumn{2}{l}
{(Luminosity constrained)} \\
\hline
$pp + pep$ & 8.1  & 1.7   &  6.75 & 0.11  & 5.91 + 0.01 \\
$^7$Be + CNO & -0.43  & 0.24   &  -0.25  & 0.11  & 0.52 + 0.12 \\
$^8$B & 0.00030  &  0.00004  &  0.00027 & 0.00003  & 0.00066 \\
$I$ & 101  &  18  &  85.32  & 0.34 &  85.31 \\
$\chi^2$ &   &    & 0.8 & & \\
Probability & 4\%  &  &  1.7\%  &  &  \\
\hline
$^a$Ref.\protect\cite{BP} \\
\end{tabular}
\end{table}

\narrowtext

\begin{table}
\caption{Differential coefficients for the flux  $\Phi_{7+}$  (10$^{10}
\nu$ cm$^{-2}$
s$^{-1}$). ($f_{CNO} = 0.185$, $f_{pep} = 0.0023$)}
\label{diff}
%\medskip
%\centering
\begin{tabular}{lrrl}
Parameter & $\frac{\partial\Phi_{7+}}{\partial X} X$	&
$\frac{\partial\Phi_{7+}}{\partial X} X$	   & Uncertainty $\frac{\Delta
X}{X}$  \\
& (Luminosity & (Luminosity & \\
X & unconstrained) & constrained) &  \% \\
\hline
$a_{C1}$ & -0.09 & -0.02 & 1.2 \\
$a_{C7}$ & +0.29 & +0.02 & 1.2 \\
$a_{C8}$ & -1.01 & -0.17 & 3 \\
$a_{G1}$ & +0.08 & -0.73 & 1 \\
$a_{G7}$ & -0.02 & +0.18 & 13 \\
$a_{G8}$ & +0.01 & -0.07 & $^{+28}_{-15}$ \\
$R_{Cl}$ & +0.79 & +0.18 & 10 \\
$R_{Kam}$ & -1.00 & -0.31 & 14 \\
$R_{Ga}$ & -0.07 & +0.65 & 11 \\
$f_{pep}$ & -0.08 & -0.06 & 10 \\
$f_{CNO}$ & +0.10 & +0.04 &  \\
$I$ & - & -0.72 & 0.4 \\
\hline
$\Phi_{7+} =$ & -0.29(16) & -0.19(8) &  \\
\end{tabular}
\end{table}


\begin{references}
\bibitem{FN1}  The data are assumed to be normally
distributed, although there is some evidence that the distributions are not
exactly normal from the asymmetric errors quoted by the experimental groups.
\end{references}
\end{document}